\documentclass[aps,onecolumn,nofootinbib,floatfix,letterpaper,showpacs,notitlepage]{revtex4-1}

\usepackage{times}
\usepackage{graphicx}
\usepackage{epsfig}
\usepackage{subfigure}
\usepackage{epstopdf}
\usepackage{amssymb,amsmath}
\usepackage{hyperref}
\usepackage{setspace}
\usepackage{url}

\usepackage[T1]{fontenc}

\hypersetup{
    colorlinks = true,
    linkcolor = blue,
    anchorcolor = blue,
    citecolor = blue,
    filecolor = blue,
    pagecolor = blue,
    urlcolor = blue
}

\begin{document}

\title{Anisotropic to Isotropic Phase Transitions in the Early Universe}

\author{Muhammad Adeel Ajaib\footnote{email: adeel@udel.edu}}
\affiliation{Department of Physics and Astronomy\\
University of Delaware, Newark, Delaware 19716.}

\begin{abstract}

We propose that the early Universe was not Lorentz symmetric and that a gradual transition to the Lorentz symmetric phase occurred. An underlying form of the Dirac equation hints to such a transition for fermions. Fermions were coupled to space-time in a non-trivial manner such that they were massless in the Lorentz violating phase. The partition function is used as a transfer matrix to model this transition on a two level thermodynamics system that describes how such a transition might have occurred. The system that models this transition evolves,  with temperature, from a state of large to negligible entropy and this is interpreted as describing the transition to a state with Lorentz symmetry. In addition to this, analogy is created with the properties of this system to describe how the fields were massless and how a baryon asymmetry can be generated in this model.

\end{abstract}

\maketitle

\section{Introduction}
The idea that the Universe is homogeneous, isotropic and that space-time is Lorentz invariant are important pillars of theoretical physics. Whereas the cosmological principal assumes the Universe to be homogeneous and isotropic, Lorentz invariance is required to be a symmetry of any relativistic quantum field theory. These requirements have robust footings, but there can possibly be scenarios where these  ideas are not sufficient to describe the dynamics of a system. Temperature fluctuations in the Cosmic Microwave Background (CMB) radiation indicate that the assumptions made by the cosmological principal are not perfect. There is no conclusive evidence of Lorentz violation to date but this  has been a topic of considerable interest and the Standard Model Extension (SME) has been constructed which includes various terms that preserve observer Lorentz transformations but violate particle Lorentz transformations \cite{Colladay:1998fq}.
 Limits have been placed on the coefficients of various terms in the SME as well \cite{Kostelecky:2008ts}. Another important question is the matter-antimatter asymmetry of the Universe which is not completely resolved. Sakharov, in 1967 derived three conditions (baryon violation, C and CP violation and out of thermal equilibrium) for a theory to satisfy in order to explain the baryon asymmetry of the Universe.
 
 Origin of fermion masses is also one of the most intriguing questions which is now close to be answered by the ATLAS and CMS experiments at the Large Hadron Collider. Hints of the Higgs boson have been seen and we will know for sure soon whether it exists or not. The formalism presented in this article might also help us answer these two important questions, namely, the baryon asymmetry of the Universe and the origin of fermion masses.
 
 We intend here to describe the evolution of a theory that violates Lorentz invariance to a theory that preserves it. The fields that are involved in the Lorentz violating theory can be viewed in analogy with fields traveling in an anisotropic medium. When the system evolves from the anisotropic to isotropic phase the symmetry of the theory is restored and the partition function formalism can be used to  better understand how this transition takes place. This formalism, we propose, can help explain the matter-antimatter asymmetry of the Universe.
 
The paper is organized as follows: In section \ref{transformations} and \ref{visualize}, we describe these transformations and propose a way to interpret them as plane wave transitions into anisotropic media. In section \ref{partition}, the partition function is used to get a better insight into how the transformations in section \ref{transformations} occur. Section \ref{interaction} illustrates how some interaction terms lead to Lorentz violating operators which are suppressed due to the transition of the system to a Lorentz symmetric phase. We conclude in section  \ref{conclusion}.
\section{Transformations leading to Covariant Dirac equation}\label{transformations}
In this section we outline a set of transformations that lead to the Dirac equation for a QED (Quantum Electrodynamics) like theory with no interaction terms. The interaction terms will be discussed in section \ref{interaction}. We start with a Dirac-like equation which involves four massless fields ($\chi_a,\chi_b,\chi_c,\chi_d$). These fields can be redefined in a simple way such that the covariant form of the Dirac equation is restored along with a mass term. In this section we will just consider the kinetic terms for the fields in the underlying theory so as to get the free Dirac equation in covariant form. 
 If we start with the following equation ($\hbar=c=1$):
\begin{eqnarray}
i \bar{\chi_a} \gamma^0 \partial_0 \chi_a + i\bar{\chi_b} \gamma^1 \partial_1 \chi_b
+i\bar{\chi_c} \gamma^2 \partial_2 \chi_c
+i\bar{\chi_d} \gamma^3 \partial_3 \chi_d
=0,
\label{eq1}
\end{eqnarray}
and transform each of the $\chi$ fields in the following manner,
\begin{eqnarray}
\chi_a(x) \rightarrow e^{i\alpha m \gamma^{0} x_{0}} \psi(x) \nonumber ,\ \
\chi_b(x) \rightarrow e^{i\beta m \gamma^{1} x_{1}} \psi(x)\nonumber \\
\chi_c(x) \rightarrow e^{i\delta m \gamma^{2} x_{2}} \psi(x)  , \ \
\chi_d(x) \rightarrow e^{i\sigma m \gamma^{3} x_{3}} \psi(x),
\label{trans1}
\end{eqnarray}
we get the  Dirac equation in covariant form, along with a mass term (using, for e.g., $e^{i\beta m \gamma^{1} x_{1}} \gamma_0=\gamma_0 e^{-i\beta m \gamma^{1} x_{1}} $),
\begin{eqnarray}
\overline{\psi}(i \gamma^{\mu} \partial_{\mu}-(\alpha+\beta+\delta+\sigma)m)\psi =0,
\label{eq2}
\end{eqnarray} 
where $\alpha,\ \beta,\ \delta \ \text{and}\ \sigma$ are real positive constants. For plane wave solution for particles, $\psi=e^{-i p.x} u(p)$, the above redefinition for the field $\chi_a$, for example, is a solution of the following  equation:
\begin{eqnarray}
\frac{\partial}{\partial t}\chi_a(x)= -i( E- \alpha  m \gamma_{0}) \chi_a(x)  ,\ \
\label{trans2}
\end{eqnarray}
with similar equations for the other fields. Equation (\ref{trans2}), is similar to equation (27) in reference \cite{braga} which is a solution of the differential equation governing linear elastic motions in an anisotropic medium (with a constant matrix, see section III of the reference). With $\alpha=0$ the left hand side is just the Hamiltonian, with the plane wave its eigenstate. 

Note that the manner in which we can transform equation (\ref{eq1}) to (\ref{eq2}) is not unique and there are various ways to do this with  different combinations of the $\chi$ fields along with the field $\psi$. A mass term ($m \overline{\chi} \chi$) for the $\chi$ fields could have  been added to equation (\ref{eq1}), but the transformations (\ref{trans1}) can be used to eliminate it. So, if we want our resulting equation to describe a massive fermion, these fields should be massless or cannot have mass term of the form $m \overline{\chi} \chi$. This argument will  be further corroborated with the results we present in section \ref{partition}. The transformation matrices in equation (\ref{trans1})  are not all unitary, the matrix $e^{i\alpha m \gamma^{0} x_{0}}$ is unitary while the rest ($e^{i\beta m \gamma^{i} x_{i}}$) are hermitian.

 The fields in equation (\ref{eq1}) can be considered as independent degrees of freedom satisfying equation (\ref{trans2}) in an underlying theory that violates Lorentz invariance. The transformations (\ref{trans1}) can, therefore,  be seen as reducing the degrees of freedom of the theory from four to one. In such an underlying theory, various interaction terms can be written for these fields. Since we intend to obtain the free Dirac equation, we have considered only kinetic terms involving the fields $\chi$.  A quadratic term involving different $\chi$ fields ($m \overline{\chi}_i \chi_j$) can be added to equation (\ref{eq1}) but this leads to a term that violates Lorentz invariance in the resulting Dirac equation. A quartic term ($c \overline{\chi}_i \chi_i  \overline{\chi}_j \chi_j$) is possible and would result in a dimension 6 operator for the field $\psi$ with the constant $c$ suppressed by the square of a cutoff scale. So, with the restriction that the resulting Dirac equation only contains terms that are Lorentz scalars the number of terms we can write for the $\chi$ fields can be limited. In other words we impose Lorentz symmetry in the resulting equation so that various  terms vanish or have very small coefficients. The interaction terms are further discussed in section \ref{interaction} of the paper.

\section{Visualizing field Redefinitions} \label{visualize}

We can visualize a global and a local transformation as transitions of plane waves to different types of media.  The wave function of a particle, for example, which comes across a potential barrier $(E>V)$ of a finite width and height undergoes a phase rotation ($e^{i k l} \psi$) upon transmission. If the width of the barrier extends to infinity, the wave function can be viewed as undergoing a position dependent phase rotation ($e^{i k x} \psi$). The transformations (\ref{trans1}) can similarly be seen as a plane wave entering an anisotropic medium. A phenomenon in optics called birefringence can be used to explain why these four fields map on to the same field $\psi$. Birefringence results in a plane wave splitting into two distinct waves inside a medium having different refractive indices along different directions in a crystal. These analogies can serve as crude sketches to visualize how the transformations in equation (\ref{trans1}) can occur.

When a polarized electromagnetic plane wave enters a birefringent material, the wave splits into two distinct waves. This can be also be seen as a change in the coordinate system of the wave. Similarly, transformations (\ref{trans1}) can be seen as rotation of the field $\psi$ in spinor space or the transition of a wave in a material that splits the field $\psi$ into four distinct waves and is anisotropic. Based on the latter view, we propose a transition from an anisotropic to isotropic phase in the early Universe whereby it became Lorentz symmetric. So the form of equation (\ref{eq1}) is not a bizarre choice of basis for writing the equation but a possible form of the equation in an anisotropic space-time. The difference also comes from the interaction terms  of these fields, which we shall discuss in section \ref{interaction}. The interaction terms  of the fields $\chi$ that are due to the anisotropic character of space-time get enhanced in the anisotropic phase. As the transition takes place these terms become suppressed.  This would not be the case if we just choose a different basis to write the equation. The suppression of the anisotropy is interpreted as the reduction of entropy or increase in order of the system discussed in section \ref{partition}.

Space-time dependent field redefinitions in the usual Dirac Lagrangian result in violation of  Lorentz invariance. For example, the field redefinition $\psi \rightarrow e^{-i a^{\mu} x_{\mu}} \psi$ leads to the Lorentz violating terms in the Lagrangian \cite{Colladay:1998fq}. This particular redefinition, however, would not lead to physically observable effects for a single fermion. A transformation of this type amounts to shifting the four momentum of the field. It can also be viewed in analogy with plane waves entering another medium of a different refractive index which results in a change in the wave number of the transmitted wave. Similarly,  transformations (\ref{trans1}) can be interpreted as transitions of a wave from an anisotropic to isotropic medium or vice versa as done in the Stroh's matrix formalism  \cite{braga}.

 For plane wave solutions of $\psi$, the $\chi$ fields have propagative, exponentially decaying and increasing solutions (for example, $e^{\pm i m x}, \ e^{\pm m x}$). This wave behavior is similar to that in an anisotropic medium or a medium made of layers of anisotropic medium. The eigenvalues of the Dirac matrices being the wave numbers of these waves in this case. The coefficients in the exponent relates to how fast the wave oscillates, decays and/or increases exponentially. The transfer matrix in Stroh's formalism describe the properties of the material and in this case can possibly represent the properties of the anisotropic phase from which the transition to the isotropic phase occurs.

In the usual symmetry breaking mechanism a Higgs field acquires a vacuum expectation value (VEV) and the resulting mass term does not respect the symmetry of the underlying group. For example, in the Standard Model, due to its chiral nature, a Higgs field is introduced in order to manifest gauge invariance. Once the Higgs field acquires a VEV the mass term only respects the symmetry of the resulting group which is $\mathrm{U(1)_{EM} }$. In our case the mass term arises after  symmetry of the Dirac equation is restored. Consider the simple case where we have one field $\chi_a$ in addition to the field $\psi$:
\begin{eqnarray} 
i \bar{\chi_a} \gamma^0 \partial_0 \chi_a + i\bar{\psi} \gamma^i \partial_i \psi
=0,
\label{eq3} 
\end{eqnarray}
 and this field transforms to the field $\psi$ as $\chi_a(x) \rightarrow e^{i\alpha m \gamma^{0} x_{0}} \psi(x),$ leading to the Dirac equation. In order to discuss the symmetries of the above equation let's assume that the two independent degrees of freedom are described by the above equation. Equation (\ref{eq3}) then has two independent global U(1) symmetries and the resulting equation has one. In fact, there is a list of symmetries of equation (\ref{eq3}) not possessed by (\ref{eq2}), for example invariance under local transformations, $\chi_a \rightarrow e^{i b^i \theta(x_i)} \chi'_a$ ($i,j=1,2,3$), where $b_i$ can be a constant vector, the matrix $\gamma_0$ or any matrix that commutes with $\gamma_0$ (e.g., $\sigma_{ij},\ \gamma_5 \gamma_i$). This implies invariance under global and local SO(3) transformations (rotations of the fields $\chi_a$ but not boosts). Similarly,  $\psi \rightarrow e^{i A \ \theta(t)} \psi'$ is a symmetry, where $A$ can be a constant or the matrix $i \gamma_0 \gamma_5$ which commutes with the three Dirac matrices $\gamma_i$. After the transformation $\chi_a \rightarrow e^{i m \gamma_0 t} \psi$ the equation is no more invariant under these symmetries and the SO(1,3) symmetry of the Dirac equation is restored along with a global U(1) symmetry.
\section{Partition Function as a Transfer Matrix}\label{partition}

In the early Universe, a transition from a Lorentz asymmetric to a symmetric phase could possibly induce transformations of the form (\ref{trans1}).  Let's again consider the simple example in equation (\ref{eq3}). For this case the eigenvalues of the Dirac matrix $\gamma_0$ define the wave numbers of the waves traveling in the anisotropic medium. The direction of anisotropy in this case is the temporal direction, which means that the time evolution of these waves is not like usual plane waves.
It is not straight forward to visualize the fields, the dynamics of whom are described by the anisotropy of space time, but we can use the partition function method to get a better insight into this. We can, by using this formalism, calculate the temperature at which the transformations in equation (\ref{trans1}) occur.

  We next perform a transition to a thermodynamics system by making the transformation $i t \rightarrow \beta$, where $\beta = 1/k_{B} T$ \cite{zee}. The partition function is then given by the trace of the transformation matrix $e^{im \gamma_0 t}$,
\begin{eqnarray} 
Z=\mathrm{Tr}(e^{m \beta \gamma_0 })=2 e^{\beta m} +2 e^{-\beta m}.
\label{eq5}
\end{eqnarray}
In order to represent the transition of the system with the above partition function the temporal transfer matrix should be unitary. This partition function is similar to that of a two-level system of spin 1/2 particles localized on a lattice and placed in a magnetic field with each state, in this case,  having a degeneracy of two. The lower energy state corresponding to spin parallel to the field ($E=-m,Z_1=e^{\beta m}$).  In this case the doubly degenerate states correspond to spins up and down of the particle or anti-particle.  For $N$ distinguishable particles the partition function is $Z^N$, $N$ here is the total number of particles and antiparticles of a particular species. So, we are modeling our system as being on a lattice with the spin along the field as representing a particle and spin opposite to the field representing an antiparticle. 
  
  The evolution of this system with temperature represents the time evolution of the system in equation (\ref{eq1}). In other words the partition function describes the evolution of these waves from anisotropic to isotropic phase  as the temperature decreases. 
   For a two level system the orientation of the dipole moments becomes completely random for large enough temperatures so that there is no net magnetization. In our case we can introduce another quantity, namely a gravitational dipole, which would imply that the four states  (particle/antiparticle, spin up/down) of $N$ such particles at high enough temperatures orient themselves in a way that the system is massless. This just serves as an analogy and does not mean that the masses are orientating themselves the same way as dipoles would do in space. The anisotropic character can be seen as mimicking the behavior of the field in a two level system. The population of a particular energy level is given by,
\begin{eqnarray}
n_{p(\overline{p})}= \frac{N e^{\pm \beta m} }{e^{\beta m} + e^{-\beta m}}.
\label{eq5b}
\end{eqnarray}
  Which shows that the number density of particles and antiparticles vary in a different way with respect to temperature.  In the early Universe, therefore the anisotropic character of space-time seems to play an important role such that particles and anti-particles behave in different manners. As the temperature decreases the number density of the anti-particles  decreases and is vanishingly small for small temperatures ($\sim e^{-2\beta m}$). 
When the decoupling temperature is attained there is a difference in the number density of the particles and antiparticles as described by equation (\ref{eq5b}). This leads to an excess of particles over antiparticles. The decoupling temperature of a particular species of particle with mass $m$ and which is non-relativistic is given by, $k_B T \lesssim  2 m$. Below this temperature the particles annihilate to photons but the photons do not have enough energy to produce the pair. This can be used to get the ratio of antiparticles over particles (matter radiation decoupling). For $\beta m \approx 0.5$, we get,
\begin{eqnarray}
\frac{n_p-n_{\overline{p}}}{n_{p}}\approx 0.6 \ .
\label{eq5c}
\end{eqnarray}
\begin{figure}
\begin{center}
\includegraphics[scale=1.1]{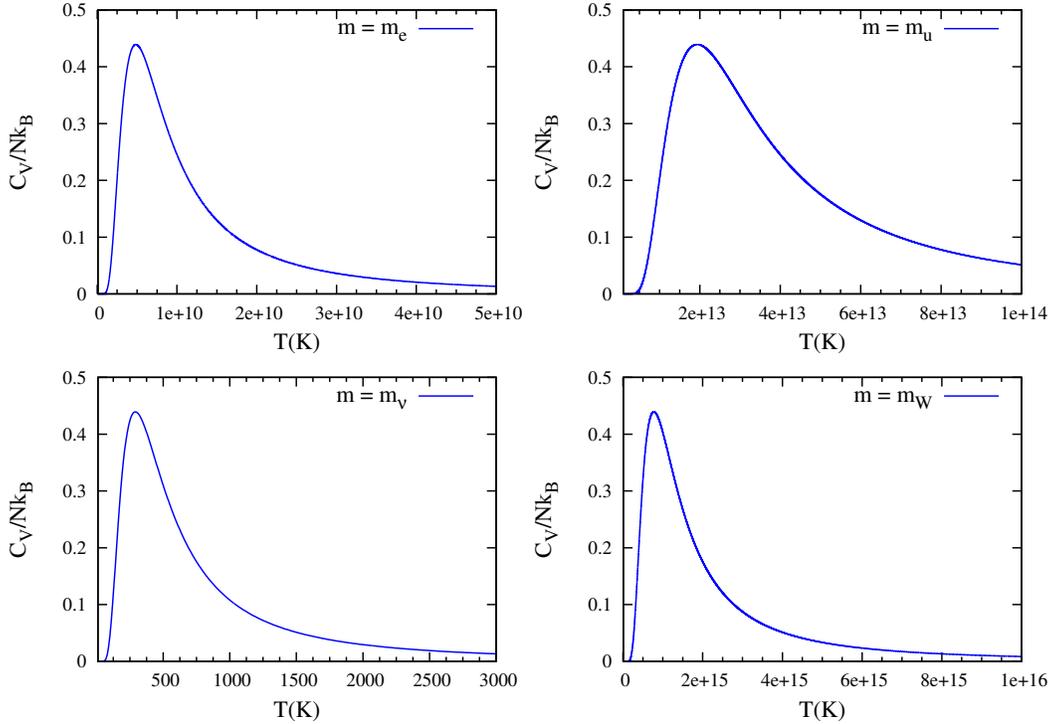}

\caption{Plot of heat capacity $C_V$ for the mass of electron, up quark, neutrino and W boson. The maximum of the heat capacity of the electron occurs at $4.8 \times 10^9 \mathrm{K}$, for the up quarks is $1.9 \times 10^{13} \mathrm{K}$, for neutrinos is $291 \mathrm{K}$ and for the W bosons is $7.8 \times 10^{14} \mathrm{K}$. We use $k_B=8.6 \times 10^{-5} \mathrm{\ eV}/\mathrm{K}$ and $m_{\nu}=0.03 \mathrm{\ eV}$.
\label{fig1}}
\end{center}
\end{figure}
Which implies an excess of particles over antiparticles and thus can serve as another possible way to explain the matter antimatter asymmetry of the Universe. This number is  very large compared to the one predicted by standard cosmology ($\sim 10^{-9}$). The above expression yields this order for $\beta m \approx 10^{-9}$ which implies a large temperature. For electrons this would imply a temperature of the order $10^{18} $K which is large and the electrons are relativistic. So if we assume that the decoupling takes place at a higher temperature, the baryon asymmetry can be explained. Even without this assumption the  conditions proposed by Sakharov can also enhance the number of particles over the antiparticles. Sakharov's conditions involve the interaction dynamics of the fields in the early Universe whereas in our case the statistical system serves more as a model describing the dynamics of space-time to a more ordered phase. 

Statistical mechanics, therefore, enables us to visualize this transition in a rather lucid way. In a two level system the net magnetization at any given temperature is analogous to the excess of particles over antiparticles in the early Universe.
The time evolution of this anisotropic to isotropic transition is modeled on the evolution of a statistical thermodynamics system with particles on a lattice placed in a magnetic field. The particles on the lattice are localized, static and have no mutual interaction. The free energy of the system is given by:
\begin{eqnarray} 
F= -N k_B T \mathrm{\ ln \{4 \mathrm{cosh}\left[m \beta \right]\} }
\label{eq6}
\end{eqnarray}
From this we can calculate the entropy $S$, heat capacity $C_V$ and mean energy ${U}$ of the system:
\begin{eqnarray}
S &=& - \left(\frac{\partial F}{\partial T}\right)_V \nonumber \\
  &=& N {k_B} \text{ln}\left \{4 \ \mathrm{cosh}\left[m \beta \right]\right \}-m k_B \beta \ \text{tanh}\left[m \beta \right] \label{eq8} \\
%
%
U &=& F+TS
 = -N m \ \text{tanh}\left[m \beta \right] \label{eq9} \\
%
%
%
C_V &=& \left(\frac{\partial U}{\partial T} \right)_V 
= N k_B m^2 \ \beta^2 \ \text{sech}^2\left[m \beta \right]
\label{eq10}
\end{eqnarray}
In Fig.\ref{fig1}, the peaks in the heat capacity represent phase transition of a particular particle species. These are second order phase transitions and the peak in the heat capacity is usually referred to as the Schottky anomaly \cite{Pathria}. Note that the phase transition we model our system on is a magnetic one. So, modeling the complex system in the early Universe on a lattice with spin 1/2 particles can reduce the complications of the actual system by a considerable amount.

The Schottky anomaly of such a magnetic system, therefore, represents phase transitions in the early Universe. For a particular species of particles the Schottky anomaly shows a peak around $mc^2\approx kT$. The phase transition for the electrons occurs at the temperature where nuclei start forming in the early Universe. For the quarks the transition temperature refers to confinement into protons and neutrons. Similarly,  W boson's transition occurs at the electroweak breaking scale. The W boson, being a spin 1 particle, is not described by the Dirac equation, but the heat capacity entails this feature of showing a phase transition for 
the energy scale relevant to the mass of a particle.

The curve for neutrinos implies that the transition temperature for neutrinos is around 291 K, which means that the density of antineutrinos  from the big bang for present neutrino background temperatures ($\sim$ 2 K) is not negligible. The ratio of antineutrinos over neutrinos for $T=2 \ \mathrm{K}$, is $n_{\overline{\nu}}/n_{\nu}= e^{-2\beta m_{\nu}}\sim  10^{-15000}$ ($m_{\nu}=2 \ \mathrm{eV}$). For an even smaller neutrino mass, $m_{\nu}=1 \times 10^{-4} \ \mathrm{eV}$, the ratio is $n_{\overline{\nu}}/n_{\nu} \sim  0.3$, which for other more massive particles is much smaller. A cosmic neutrino and antineutrino background is one of the predictions of standard cosmology but is still unobserved. This model predicts  an antineutrino background much less than the neutrino one. 

\begin{figure}
\begin{center}
\includegraphics[clip=true,trim=0mm 23mm 0mm 23mm,scale=1.3]{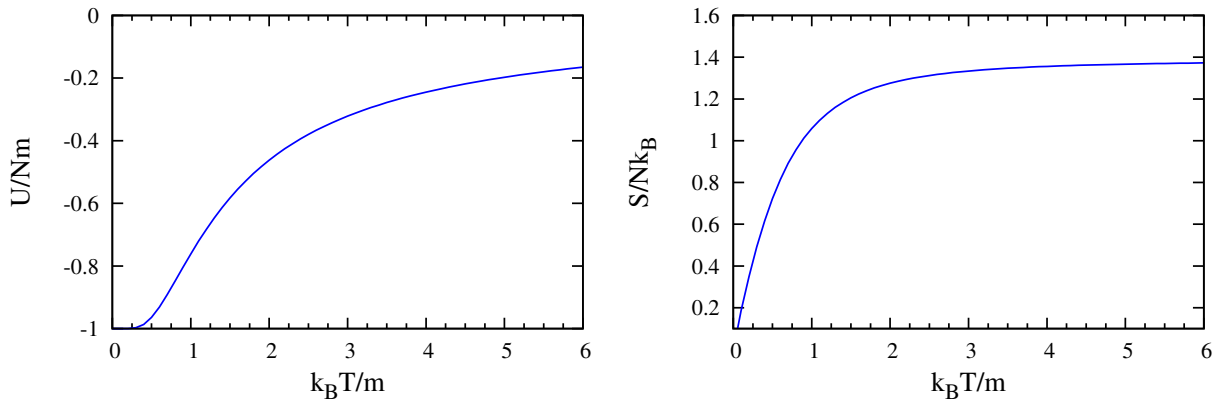}

\caption{Plot of entropy and energy for a particle of mass m. For large enough temperatures the energy of the system approaches zero and the entropy approaches the limiting value of $N k_B \mathrm{ln 4}$.
\label{fig2}}
\end{center}
\end{figure}

 In Fig.\ref{fig2}, the plots of mean energy and entropy are shown in dimensionless units. In  the massless limit for fermions, the entropy attains its maximum value of $N k_B \mathrm{ln 4}$. The plots show that the energy of the system approaches zero as the temperature approaches infinity. This situation is analogous to the spins being completely random at high temperatures for the two level system. The same way that the magnetic energy of the system on the lattice is zero at high temperatures, the mass of this system is zero in the very early Universe. As the temperature decreases the energy of the  system attains it minimum value ($U=-Nm$) and the particles become massive at the temperature less than the value given by the peak of the heat capacity. The entropy for high temperatures asymptotically approaches its maximum value of $N k_B \mathrm{ln 4}$. 
 
 The value of the parameter $\mathrm{k_B T/m}$ at the peak of the heat capacity curves gives the temperature at which the transition takes place for a particular species. The transition for each field, therefore, depends on the energy scale relevant to its mass. So, this means that each field was experiencing the anisotropy of space time in a different manner whereas space-time itself was expanding towards a Lorentz symmetric phase. This can be understood if we imagine a material with anisotropies. Plane waves of different wavelengths inside the material experience the anisotropies in different ways. So the analogy discussed in section \ref{visualize} tells us that the more massive the field, the faster it will oscillate, exponentially increase or decrease. A less massive field having a larger wavelength ($\lambda= \hbar/mc$) therefore would experience the anisotropies when their scale is much larger.

According to the statistical thermodynamics model that describes this transition, as this phase transition occurs antiparticles will start changing into particles and as can be seen from the figure the system will move towards all spins aligned parallel with the ``field", i.e., towards being particles. From Fig.\ref{fig2} we can see that the energy of the system starts attaining the minimum value as the temperature decreases where all particles are aligned with the field and are ``particles". The plot of entropy vs. temperature also represents an important feature of these transformations. The entropy decreases with decreasing temperature and this represents the transition to a more ordered phase using equations  (\ref{trans1}). The plots of energy of the system U in Fig.\ref{fig2} show that the system will eventually settle down to the lowest energy state which in this case means that the system will have almost all particles with negligible number of antiparticles. In short, the plot of the heat capacity reflects the phase transitions, the plot of  energy U represents the transition from massless to massive states and the plot of entropy represents the transition of space time to a more ordered phase.
%
%

 The Big Bang theory is one of the most promising candidates to describe how the Universe began. According to this theory, the Universe expanded from a singularity where curved space-time, being locally Minkowskian, eventually became flat. It is possible that there even was  a transition to the Minkowski space from a non-Minkowski one. If the Universe began with a state of maximum entropy than we can very well assume that space-time was not Minkowskian even locally. The fields that dwell in space-time are representation of the symmetry group that describes it. The $\chi$ fields in the underlying theory, described by equation (\ref{eq1}), are therefore, not representations of the Lorentz group. The CPT theorem assumes symmetries of Minkowski space-time in implying the similarities between particles and antiparticles. If the underlying theory is not Minkowskian than particles and antiparticles can behave differently and this is what the model described in this section implies.

The occurrence of the Schottky anomaly has motivated the study of negative temperatures \cite{Ramsey}. Note that the partition functions is invariant under the transformation $T \rightarrow -T$ but the equations for the free energy, entropy and energy are not. The existence of negative temperatures has been observed in experiments. Negative temperatures, for example, can be realized in a system of spins if the direction of the magnetic field is suddenly reversed for a system of spins initially aligned with the magnetic field \cite{Pathria}.  Similarly, as described in reference \cite{Ramsey} the allowed states of the system must have an upper limit. Whereas this is not the case for the actual particles in the early Universe, the statistical mechanics system on which it can be modeled on has this property. A negative temperature system would eventually settle down to the lower energy state ($U=Nm$) which in our case would mean that the Universe would ends up having more antiparticles than particles. This is yet another interesting insight we get by modeling the early Universe on a two state system.

 The system on the lattice eventually has only spins aligned with the field and which are interpreted as particles. This means that the partition function can be written as,
\begin{eqnarray}
Z=e^{\beta m}=e^{(-\beta) (-m)} \nonumber .
\end{eqnarray}
Which implies that the system can eventually  be seen as spins aligned towards the field evolving in a positive temperature system or spins aligned opposite to the field evolving in a negative temperature system. This gives a correlation of particles with positive temperature systems and antiparticles with negative temperature systems. This view in a way coincides with the fact that the resulting space-time configuration describes both particles and antiparticles in a similar but independent way.

We can also notice that in the statistical system modeled for the transition, the reason due to which entropy of the system decreases is the presence of an external field. As in a paramagnet, the spins remain randomly align in the absence of a magnetic field. In the system considered, spins interact with the magnetic field and not with each other. It is difficult to say what played the role of this field in the early Universe, but since the field is the reason entropy reduces, there should be a physical quantity present in the early Universe playing an analogous role. This quantity, for instance, can be speculated to be the expansion of the Universe.
\section{Interaction terms and the SME} \label{interaction}
 In section \ref{transformations} we only considered terms  in the underlying theory that lead to the free Dirac equation. In this section we shall include some interaction terms and see how they lead to LV operators one of which is considered in the SME. Since we do not have Lorentz symmetry, the underlying theory can have a large number of terms. We will therefore restrict ourselves to interaction terms that lead to dimension 4 operators and only involve the gamma matrices. Also, the form of the Dirac equation restricts the number of ways we can start with an underlying LV theory. 
 In the following we add a few interaction terms to equation (\ref{eq1}): 
\begin{eqnarray}
i  \bar{\chi_a} \gamma^0 \partial_0 \chi_a + i  \bar{\chi_b} \gamma^1 \partial_1 \chi_b
+i    \bar{\chi_c} \gamma^2 \partial_2 \chi_c
+i   \bar{\chi_d} \gamma^3 \partial_3 \chi_d + I^{int}_1 + I^{int}_2
=0,
\label{eq2b}
\end{eqnarray}
where $I^{int}_1$ and $I^{int}_2$ include operators that lead to dimension 4 LV operators after the transformations,
\begin{eqnarray}
 I^{int}_1 &=&  c_1 \overline{\chi}_a \gamma^0 \chi_a a_0 
+ c_3 \overline{\chi}_b \gamma^1 \chi_b a_1 
+ c_5 \overline{\chi}_c \gamma^2 \chi_c a_2 
+ c_7 \overline{\chi}_d \gamma^3 \chi_d a_3  \nonumber  \\
\vspace{2mm}
&+& c_{9} \overline{\chi}_a \gamma^0 \gamma_5 \chi_a b_0 
+ c_{11} \overline{\chi}_b \gamma^1 \gamma_5 \chi_b b_1 
+ c_{13} \overline{\chi}_c \gamma^2 \gamma_5 \chi_c b_2 
+ c_{15} \overline{\chi}_d \gamma^3 \gamma_5 \chi_d b_3+.... \ , \nonumber
\label{eq2c}
\end{eqnarray}
\begin{eqnarray}
 I^{int}_2 &=& c_{2} m \overline{\chi}_a \chi_b + c_6 \overline{\chi}_a \gamma^0 \chi_b a_0 
+ c_4 \overline{\chi}_b \gamma^1 \chi_c a_1 
+ c_6 \overline{\chi}_c \gamma^2 \chi_d a_2 
+ c_8 \overline{\chi}_d \gamma^3 \chi_a a_3+... \ , \nonumber 
\vspace{2mm}
%
%
%
\label{eq2d}
\end{eqnarray}
%
%
%
where $I^{int}_1$ includes interaction terms of the same fields and $I^{int}_2$ involve different fields. As the system transits to a more ordered phase using (\ref{trans1}) the coefficients of the terms leading to LV operators become suppressed. For example, after the transformations, the terms in $I^{int}_1$ would lead to the term in the SME with coefficient $a_{\mu}$. The term $\overline{\psi}\gamma^{\mu}\psi a_{\mu}$ in the SME do not have any physical effects for a single fermion since the term can be canceled with a redefinition of the field $\psi \rightarrow e^{i a^{\mu}x_{\mu}}\psi$, but can have physical effects if the theory has more than one fermions. Note that the SME term $\overline{\psi}\gamma^{\mu}\gamma_5\psi b_{\mu}$ with a constant $b_{\mu}$ is not generated from $I^{int}_1$. 

The coefficients $c_i$ are dimensionless quantities which measure the anisotropies of the system and therefore can be taken to be proportional to the entropy $\mathrm{S/k_B}$ of the system in equation (\ref{eq8}). As the system moves towards a Lorentz symmetric phase the entropy approaches zero and therefore the coefficients become vanishingly small. We can also impose four global U(1) symmetries in equation (\ref{eq2b}) and this would forbid the terms in $I^{int}_2$. The symmetries can then be gauged to get interactions of each of the $\chi$ fields with each component of the gauge field (e.g. $\overline{\chi}_a \gamma^0 \chi_a A_0(x)$) and this would convert to the usual gauge interaction term in the resulting covariant Dirac equation.

As mentioned earlier, the properties of the gamma matrices and the form of the Dirac equation restricts the ways we can start with an underlying LV theory. The gamma matrices commute with only the identity matrix and anti-commute with $\gamma_5$. The other possible way to start with an underlying theory is to use the $\gamma_5$ matrix. If we just include the $\gamma_5$ matrix with the gamma matrices in (\ref{eq1}) than we do not retrieve the Dirac equation. This is also the case when the $\gamma_5$ matrix is included in the transformations. However, if we start with both a $\gamma_5$ in the equation (e.g. $\overline{\chi}_a \gamma^0 \gamma_5 \partial_0\chi_a $) and perform the transformations (\ref{trans1}) with a $\gamma_5$ (e.g. $\chi_a \rightarrow e^{i m \gamma^0 \gamma_5 t} \psi$) than the Dirac equation is retrieved with a $\gamma_5$ and this can be rotated away with a redefinition of the field $\psi$ in the free Dirac equation. Since the temporal transfer matrix is not unitary in this case, the analogy made in section \ref{partition} will not apply and we get an oscillating partition function. This would also effect the resulting LV terms we get in the SME. In this case we get a term $\overline{\psi}\gamma^{\mu}\gamma_5\psi b_{\mu}$ with a constant four vector $b_{\mu}$, whereas $a_{\mu}$ is not constant in  $\overline{\psi}\gamma^{\mu}\psi a_{\mu}$. 
\section{Conclusions}\label{conclusion}
Considering that the Dirac equation can be written in the form of an underlying theory that violates Lorentz invariance, we suggest that such a transition took place for fermions in the early Universe. We propose that space-time was not Lorentz symmetric and that a gradual transition to the Lorentz symmetric phase occurred. The fields in the underlying Lorentz violating theory are massless and transformations were performed that restore the Dirac equation to its covariant form along with a mass term for the fermions.  

The underlying theory depicting the Lorentz violating phase has interaction terms of the fields. As the transition takes place, these interaction terms result in suppressed Lorentz violating terms some of which can be identified with terms in the SME (Standard Model Extension).  The partition function formalism is then used to model these transformations on the evolution of a system of spin 1/2 particles on a lattice placed in a magnetic field. Symmetry breaking in this case takes place on this lattice, whereas, it is restored in the Dirac equation. The transition to the Lorentz symmetric phase in the early Universe can be modeled on this thermodynamic system. 

The behavior of  the fields in the anisotropic phase is suggested to be similar to that of plane waves in anisotropic media. The eigenvalues of the transfer matrices give the wavenumbers of the waves in the anisotropic media. The wavelengths given by the temporal transfer matrix ($\hbar / mc$) show that each fermion field experienced the anisotropy of space-time in different ways. The reason entropy decreases in the statistical system is the presence of an external magnetic field. The expansion of the Universe is interpreted as playing this role. The fields with small masses, and hence large wavelengths, undergo phase transitions later as the scale of the anisotropies get larger with the expansion.

   We showed that modeling the transition in such a manner can describe three important features of the early Universe: (1) The heat capacity shows occurrence of  phase transitions. (2) The mean energy of the system shows how the particles became massive from being massless. (3) The plot of entropy depicts the occurrence of a transition to a more ordered phase interpreted as the Lorentz symmetric phase. At any given temperature the net magnetization measures the excess of particles over antiparticles. We suggest that this model can be used to explain the matter antimatter asymmetry of the Universe. Also, since space-time is not Minkowskian in the underlying theory, the CPT theorem does not hold, implying a difference in the behavior of particles and antiparticles. This is in agreement with the analogy created with the statistical system whereby spin up and down particles behave in different ways with the evolution of the system. This formalism can arguably serve as another possible way to explain the origin of fermion masses till the final results  related to the Higgs boson are presented.

\section{Acknowledgements}
The author would like to express his deep gratitude to Alan Kostelecky and Dmitry Gorbunov for very fruitful discussions and suggestions. I would also like to thank Fariha Nasir, Hassnain Jaffari, Ilia Gogoladze and Matthew DeCamp for useful discussions and comments.

\end{document}